\documentclass[letter]{jpsj3}
\usepackage{txfonts}

\usepackage{amssymb,epsfig,latexsym}
\usepackage{amsmath,amscd}
\usepackage{graphicx}
\usepackage[normalem]{ulem}
\usepackage{color}

\title{Quiescent Light Curve of Accreting Neutron Star MAXI J0556-332}
\author{
    Helei Liu$^{1,2}$\thanks{l.helei@mail.scut.edu.cn},
    Yasuhide Matsuo$^1$,
    Masa-aki Hashimoto$^1$,
    Tsuneo Noda$^3$,
    Masayuki Y. Fujimoto$^4$
}
\inst{
  $^1$Department of Physics, Kyushu University, Fukuoka 812-8581, Japan \\
  $^2$Department of Physics, Xinjiang University, Urumqi 830046, China \\
  $^3$Kurume Institute of Technology, Fukuoka 830-0052, Japan \\
  $^4$Department of Physics, Hokkaido University, Sapporo 060-8810, Japan
}

\abst{
  MAXI J0556-332 is the hottest transient accreting neutron star at the beginning of its quiescence.
  A theoretical model with crustal heating indicates that
    an additional shallow heat source of ~$Q_{\rm shallow} > 6$ MeV per accreted nucleon is required in the shallow outer crust with respect to the deeper star crust
    by considering the observed decline in accretion rate at the end of outburst. However, the physical source of this shallow heating is still unclear.
  In the present investigation,
    we performed stellar evolutionary calculations, adopting the effects of outburst behavior of the accretion rate.
  As a consequence,
    we find that the quiescent light curve of MAXI J0556-332 can be well explained by the nuclear energy generation due to the hot CNO cycle.
}

\begin{document}

\maketitle

%%%%%%%%%%%%%%%%%%%%%%%%%%%%%%%%%%%%%%%%%%%%%%%%%%%%%%%%%%%%%%%%%%%%%%%%
%%%% Section I  %%%%%%%%%%%%%%%%%%%%%%%%%%%%%%%%%%%%%%%%%%%%%%%%%%%%%%%%%%%
%%%%%%%%%%%%%%%%%%%%%%%%%%%%%%%%%%%%%%%%%%%%%%%%%%%%%%%%%%%%%%%%%%%%%%%%
%\section{Introduction}

The observations of quiescent X-ray luminosity from accreting neutron star transients has opened a new probe for exploring the physics of the neutron star structure~\cite{2007PhR...442..109L,2007ApJ...660.1424H,2009ApJ...691.1035H,2015MNRAS.447.1598B}. 
A transiently accreting neutron star experiences periods of outburst activity separated by long phases of relative quiescence periods, during which accretion is switched off or strongly suppressed. 
The accreted crust is heated during outburst by electron captures, neutron emission, and pycnonuclear reactions that release 1$-$2 MeV per accreted nucleon~\cite{1990A&A...227..431H,2003A&A...404L..33H,2008A&A...480..459H}.
The energy release due to crustal heating heats the star enough to produce the quiescent light curve which is consistent with the observations~\cite{2003A&A...407..265Y,2015MNRAS.447.1598B,2016arXiv161009100M}.

However, it has been advocated that the shallow outer crust ($\rho\lesssim10^{10}\rm ~g~cm^{-3}$) should be heated with respect to the deeper neutron star crust to explain the temperatures observed in the first months of relaxation for several sources, for example, the light curves of KS1731-260 and MXB 1659-29 required a shallow heat source of $\approx1~\rm MeV$ in the calculation by Brown \& Cumming~\cite{2009ApJ...698.1020B}.

MAXI J0556-332 is such an accreting neutron star transient; it was discovered in January 2011 through MAXI \cite{2011ATel.3102....1M} and went into quiescence in May 2012. After the outburst period had continued for more than 16 months,  it returned to quiescence \cite{2014ApJ...795..131H}.
The observations of MAXI J0556-332 through Chandra and XMM-Newton were initiated and had been analyzed using a variety of X-ray instruments. As a consequence, the Swift/XRT light curve indicates that there is an exponential decay time scale of $\sim$ 3.3 days for the last $\approx14$ days of the outburst \cite{2014ApJ...795..131H}. The data of MAXI J0556-332 obtained through the Rossi X-ray Timing Explorer (RXTE) show similarities to the class of low mass X-ray binaries known as ``Z-sources"~\cite{2014ApJ...795..131H,2013PASJ...65...58S}, implying that the neutron star in MAXI J0556-332 accretes at near- or super-Eddington limit.

Furthermore, this star has been thought to be the hottest quiescent neutron star in this class, in which crustal heating models cannot
explain the observational data of light curve. In Deibel's detailed calculation \cite{2015ApJ...809L..31D}, an additional shallow heat source $Q_{\rm shallow}\approx6-16~\rm MeV$ per accreted nucleon is required by considering of the decay of the accreting rate at the end of the outburst. However, the physical source of this shallow heating is still unknown.
Furthermore, we cannot adopt the relation between the photospheric temperature and the temperature at the bottom of accreted envelop
if some heatings occur in the envelope.
Regarding the crustal heating, the heat flow from the crust to the inner regions has been shown to be
important for explaining the observation related to the X-ray burst due to helium shell burning~\cite{2015ApJ...809L..31D}.

In the present paper, we present theoretical fits to the light curve for the observational data of  the transient source MAXI J0556-332 by adopting the stellar evolutionary code with the effect of the outburst behavior; the accretion rate does not turn off instantaneously at the end of the outburst.
We assume a similar decay time scale obtained from the Swift/XRT observations.
In the accretion layer, the nuclear reaction from the hot CNO cycle~\cite{1981ApJS...45..389W} is included because it will occur when the temperature increases in the range of $0.2\leq T_9\leq 0.5~$($T_9=T/10^9$~K).
During the quiescence, the energy deposited through crustal heating, compressional heating, and
 hot CNO cycle will be released gradually.

%In section II, we outline our thermal evolution model to fit the light curve of MAXI J0556-332. Our results compared with
%observations are presented in section III.  Conclusions are given in the last section.

%%%%%%%%%%%%%%%%%%%%%%%%%%%%%%%%%%%%%%%%%%%%%%%%%%%%%%%%%%%%%%%%%%%%%%%%
%%%% Section II  %%%%%%%%%%%%%%%%%%%%%%%%%%%%%%%%%%%%%%%%%%%%%%%%%%%%%%%%%%
%%%%%%%%%%%%%%%%%%%%%%%%%%%%%%%%%%%%%%%%%%%%%%%%%%%%%%%%%%%%%%%%%%%%%%%%
%\section{Cooling Model of the MAXI J0556-332}

We performed calculations of the thermal evolution of neutron stars in hydrostatic equilibrium by using the spherical symmetric stellar evolutionary code~\cite{1984ApJ...278..813F,1984PASJ...36..199H},
  which includes full general relativistic effects  formulated by Throne~\cite{1977ApJ...212..825T}.
Basic simultaneous differential equations are written as follows:
\begin{eqnarray}
  \frac{\partial M_{tr}}{\partial r} \hspace*{-2mm}& = &\hspace*{-2mm} 4\pi r^{2} \rho~, \label{eq:1} \\
  %
  %\frac{\partial P}{\partial r}\hspace*{-2mm} & = &\hspace*{-2mm} -\frac{GM_{tr}\rho}{r^{2}}
  %    \left(1+\frac{P}{\rho c^{2}}\right)
  %    \left(1+\frac{4\pi r^{3}P}{M_{tr}c^{2}}\right) \nonumber \\
  %&&
  %    \left(1-\frac{2GM_{tr}}{c^{2}r}\right)^{-1}~, \label{eq:2} \\
  %
  \frac{\partial P}{\partial r}\hspace*{-2mm} & = &\hspace*{-2mm} -\frac{GM_{tr}\rho}{r^{2}}
      \left(1+\frac{P}{\rho c^{2}}\right)
      \left(1+\frac{4\pi r^{3}P}{M_{tr}c^{2}}\right) \left(1-\frac{2GM_{tr}}{c^{2}r}\right)^{-1}~, \label{eq:2} \\
  \frac{\partial (L_{r}e^{2\phi/c^{2}})}{\partial M_{r}}\hspace*{-2mm} & =\hspace*{-2mm} &
      e^{2\phi/c^{2}}\left(\varepsilon_{\rm n}+\varepsilon_{\rm g}-\varepsilon_{\nu}
      \right)~, \label{eq:3} \\
  \frac{\partial \ln T}{\partial \ln P}\hspace*{-2mm} & =\hspace*{-2mm} & {\rm min}(\nabla_{\rm rad}, \nabla_{\rm ad})~, \label{eq:4} \\
  \frac{\partial M_{tr}}{\partial M_{r}}\hspace*{-2mm} & =\hspace*{-2mm} & \frac{\rho}{\rho_0}
  \left(1-\frac{2GM_{tr}}{c^{2}r}\right)^{1/2}~, \label{eq:5}\\
  \frac{\partial \phi}{\partial M_{tr}}\hspace*{-2mm} & =\hspace*{-2mm} & \frac{G(M_{tr}+4\pi r^{3}P/c^{2})}
    {4\pi r^{4}\rho}\left(1-\frac{2GM_{tr}}{c^{2}r}\right)^{-1}. \label{eq:6} %
\end{eqnarray}
Here, $M_{tr}$ and $M_r$ are the gravitational and rest masses, respectively, in radius $r$;
$\rho$ and $\rho_0$ denote the total mass energy and rest mass densities, respectively;
$P$ and $T$ are the pressure and local temperature, respectively;
$\varepsilon_{\rm n}$ and $\varepsilon_{\rm g}$ are the energy generation rates by nuclear burning and gravitational energy release, respectively. 
Furthermore, $\varepsilon_\nu$  represents energy loss rate by neutrino emission;
$\nabla_{\rm rad}$ and $\nabla_{\rm ad}$ are the radiative and adiabatic gradients, respectively;
$\phi$ is the gravitational potential in unit mass.

We adopt the fraction of the rest mass $q~[=M_r/M(t)]$, which is adopted when the stellar mass varies \cite{1981PThPS..70..115S,1984ApJ...278..813F}. The gravitational energy release $\varepsilon_{\rm g}$ in Eq.~(\ref{eq:3})
is expressed as $\varepsilon_{\rm g} = \varepsilon_{\rm g}^{\rm (nh)} + \varepsilon_{\rm g}^{\rm (h)}$, where each part in the right-hand
side is written as follows:
\begin{eqnarray}
    \varepsilon_{\rm g}^{\rm (nh)}\hspace*{-2mm} & = &\hspace*{-2mm} -\exp\left(-\frac{\phi}{c^2}\right)\left(T\frac{\partial s}{\partial t}\Bigg|_q + \mu_i\frac{\partial N_i}{\partial t}\Bigg|_q\right), \label{eq:7} \\
    \varepsilon_{\rm g}^{\rm (h)}\hspace*{-2mm} & = &\hspace*{-2mm} \exp\left(-\frac{\phi}{c^2}\right)\frac{\dot{M}}{M}\left(T\frac{\partial s}{\partial \ln q}\Bigg|_t+\mu_i\frac{\partial N_i}{\partial \ln q}\Bigg|_t\right), \label{eq:8} %
\end{eqnarray}
where $\mu_i$ and $N_i$ are the chemical potential and number per unit mass of the $i$-th element, respectively, and
$t$ is the Schwarzschild time coordinate. 
In Eq.~(\ref{eq:8}), $\dot{M}$  is the mass accretion rate.
Eqs.~(\ref{eq:7}) and (\ref{eq:8})  are respectively called nonhomologous and homologous terms, where the latter implies a homologous compression due to the accretion~\cite{1984ApJ...278..813F}.
Note that compressional heating due to the accretion contributes significantly to the heat source as well as nuclear burning. 

We adopt an equation of state (EoS) by Lattimer \& Swesty\cite{1991NuPhA.535..331L} with the incompressibility of 220 MeV in the inner layers~($\rho\geq10^{12.8} \rm~g~cm^{-3}$) and connect it to EoS in Ref. \citen{1971ApJ...170..299B}) for the outer layers
($\rho<10^{12.8} \rm~g~cm^{-3}$). 
The neutrino emission process is set to a slow cooling process; electron-positron pair, photo, plasmon processes~\cite{1979ApJ...232..541F,1969PhRv..180.1227F,1967ApJ...150..979B}; bremsstrahlung process; modified Urca (MURCA) process.
We do not include the neutrino emission through pion condensation.
Therefore, the dominant processes are MURCA and bremsstrahlung processes.
The corresponding energy loss rates are briefly written as follows~\cite{1979ApJ...232..541F}:
\begin{eqnarray}
  \varepsilon_{\nu}^{\rm MURCA} &=& 2.6 \times 10^{20} \left( \frac{\rho}{\rho_{\rm nuc}} \right)^{2/3} T_9^{8} ~{\rm erg~cm^{-3} s^{-1}}, \\
  \varepsilon_{\nu}^{\rm brems.}   &=& 3.8 \times 10^{19} \left( \frac{\rho}{\rho_{\rm nuc}} \right)^{1/3} T_9^{8} ~{\rm erg~cm^{-3} s^{-1}},
\end{eqnarray}
where $\rho_{\rm nuc}$ is the nuclear density.
On the other hand, the neutrino emission rates have been studied; however, the uncertainty of a factor of ten remains because of the insufficient understanding of the symmetry energy and nucleon effective mass in the dense matter \cite{Yin+2017}.
Moreover, we do not include the strong process such as pion condensation.
Although pion condensation accompanies the strong neutrino loss rates, the effects of the super-fluidity may reduce the neutrino emissions.
However, the critical temperature for the super-fluid to occur is very uncertain.
As our aim is to present a possible heat source instead of unknown source, we neglect the strong neutrino emission for simplicity.

The energy generation includes crustal heating~\cite{1990A&A...227..431H}, compressional heating~\cite{1984ApJ...278..813F}, and the hot CNO cycle~\cite{1981ApJS...45..389W}. Crustal heating has the following form:
\begin{equation}
    Q_i=6.03 \times \dot{M}_{-10}~q_i~ 10^{33}~ \rm erg~s^{-1}~, \label{eq:crustheat}
\end{equation}
where $\dot{M}_{-10}=\dot{M} / (10^{-10}~M_\odot~{\rm yr^{-1}})$ is the mass accretion rate, and $q_i$ is the deposited heat per nucleon on the $i$-th reaction layer. Detailed tables for $q_i$ can be found in Ref. \citen{1990A&A...227..431H}). The energy generation rate $\varepsilon_n$ in Eq.~(\ref{eq:3}) can be obtained from $Q_i/\delta M$, where $\delta M$ is the mass of the $i$-th reaction layer.

The other heating process is viscous heating,
which originates because of the $r$-mode instability associated with the rapidly rotating NSs \cite{Andersson1998,Lindblom+1998}.
This process may affect the temperature evolution of NSs; \cite{Levin1999}
however, we do not include it in our study because the dimensionless amplitude of the $r$-mode is very uncertain, and the heating rates is unclear \cite{Chugunov+2017}.

The accreted matter is assumed to have a uniform chemical composition with each mass fraction ($X, Y, Z) = (0.73, 0.25, 0.02)$, where $X, Y$, and $Z$ represent the mass fractions of hydrogen, helium, and heavy elements, respectively.
We adopt the simple formula of the nuclear energy generation rate for the hot CNO cycle~\cite{1981ApJS...45..389W} for the temperature range of $0.2\leq T_9\leq 0.5$:
\begin{equation}
    \varepsilon_{\rm hCNO}=5.86\times 10^{15}Z' \rm ~erg~g^{-1}~s^{-1}~, \label{eq:CNO}
\end{equation}
where $Z'$ represents the sum of the mass fractions of CNO isotopes inside the accreted envelope,
that is, $Z' = 0.02$.

%%%%%%%%%%%%%%%%%%%%%%%%%%%%%%%%%%%%%%%%%%%%%%%%%%%%%%%%%%%%%%%%%%%%%%%%
%%%% Section III  %%%%%%%%%%%%%%%%%%%%%%%%%%%%%%%%%%%%%%%%%%%%%%%%%%%%%%%%%
%%%%%%%%%%%%%%%%%%%%%%%%%%%%%%%%%%%%%%%%%%%%%%%%%%%%%%%%%%%%%%%%%%%%%%%%
%\section{Theoretical Results Compared with Observations}

We construct the initial models of a neutron star to be accreted at around the Eddington rate $dM/dt =2.73 \times 10^{-8} M_\odot~\rm yr^{-1}$ with and without the crustal and compressional heatings and the hot CNO cycle. The initial model corresponds to a steady state,
in which the nonhomologous part of the gravitational energy release Eq.~(\ref{eq:8})
can be neglected.~\cite{1984ApJ...278..813F,1984PASJ...36..199H}
The gravitational mass and radius of the neutron star are $M=1.54~M_\odot$ and $R=12.48~\rm km$, respectively.
We assume a decay time scale $\tau$ ($e$-folding time) at the beginning of cooling, where the time scale is chosen
to match the duration of the MAXI outburst~\cite{2014ApJ...795..131H}.
Futhermore, we construct the light curve
by tuning the mass of the accreted elements $\Delta M$ and  $\tau$ for the accretion rate.

\begin{figure}
    \centering
    \includegraphics [scale=0.35]{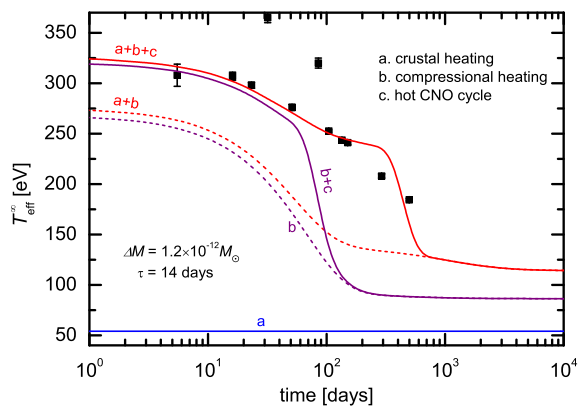}
    \caption{
        Model fit to the quiescent light curve of MAXI J0556-332 for a neutron star with $M=1.54~M_\odot$ and $R=12.48~\rm km$
        considering the hot CNO cycle. The two data with high effective temperatures above the theoretical light curve are considered to be
        contaminations from residual accretion~\cite{2015ApJ...809L..31D}.
        The observational data are taken from Ref.~(\citen{2014ApJ...795..131H}).
    }
    \label{lc}
\end{figure}

Figure~\ref{lc} shows the theoretical light curve with $\Delta M=1.2\times10^{-12}M_\odot$ and  $\tau=14~\rm days$.
Five curves are drawn involving four cases: curve `a' includes only the crustal heating~(\ref{eq:crustheat}),
and curve `b' only the compressional heating~(\ref{eq:8}). The case for the heating due to the hot CNO 
cycle~(\ref{eq:CNO}) is indicated by `c'.
Note that the time is measured from the end of the outburst~\cite{2014ApJ...795..131H}, and
our cooling curve (`a+b+c') can well reproduce the light curve as a whole.
The dotted curves (`a+b' and `b') show the light curves without the additional energy source of the hot CNO cycle.
We recognize that the nuclear energy source of the hot CNO cycle can provide significant heating
    to increase the luminosity in addition to the compressional and crustal heatings until approximately 300 days.
Both the compressional and crustal heatings maintain the heating according to the decreasing accretion rate after 500 days.
In particular, the light curve after 700 days is predicted owing to the two heat sources, as shown in Fig~\ref{lc},
    in which the contribution from the compressional heating is larger than the crustal heating by a factor of 1.8.

Note that the two observational data with high effective temperatures
    that are not fitted by our light curve are attributed to the increase in the accretion rates~\cite{2014ApJ...795..131H} or contaminations
    due to the residual accretion~\cite{2015ApJ...809L..31D}.
If we include sudden increases in accretion rates,
    the temperature in the accretion layers will increase and furthers the nuclear burning results.
However, this may result in a very complex thermal structure.

\begin{figure}
    \centering
    \includegraphics [scale=0.38]{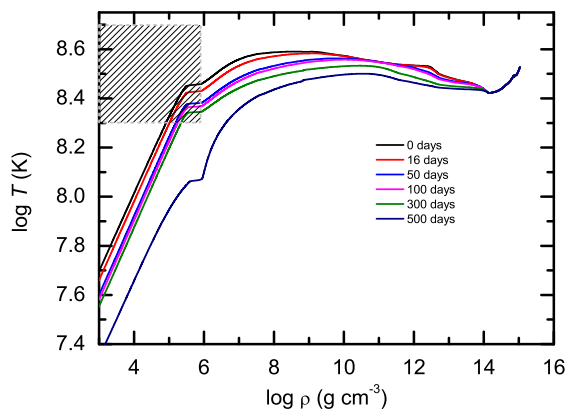}
    \caption{
        Changes in the temperature during the quiescence era against the density.
        The shadowed rectangle  indicates the region where the hot CNO cycle operates effectively.
        The right edge of the rectangle area indicates the bottom of the accreted envelope.
    }
    \label{trho}
\end{figure}

The fitting is not adequate for the last two observations
because we use the approximate nuclear energy generation rate of Eq.~(\ref{eq:CNO}) for the hot CNO cycle.
Figure \ref{trho} shows changes in the temperature distribution against the
density during the quiescence. The shadowed area indicates the temperature region for the envelope to reach the condition where the
hot CNO cycle occurs significantly. 
The hot CNO cycle can be seen to operate until 500 days before the flat shape of the quiescence appears. 
Therefore, we must at least consider the effects of changes in the abundances during the operation of the hot CNO cycle. 
Furthermore, our assumption for the exponential decay of the accretion rate may be inadequate to apply the last two observations of the light curve.
If a reasonable initial model reflecting thermal history of the previous accretions are constructed,
detailed calculations of the large nuclear reaction network could reproduce the light curve.

%%%%%%%%%%%%%%%%%%%%%%%%%%%%%%%%%%%%%%%%%%%%%%%%%%%%%%%%%%%%%%%%%%%%%%%%
%%%% Section IV  %%%%%%%%%%%%%%%%%%%%%%%%%%%%%%%%%%%%%%%%%%%%%%%%%%%%%%%%%
%%%%%%%%%%%%%%%%%%%%%%%%%%%%%%%%%%%%%%%%%%%%%%%%%%%%%%%%%%%%%%%%%%%%%%%%
%\section{Concluding Remarks}

We constructed a theoretical light curve of MAXI by using a stellar evolutionary code and tried to fit the observations.
We included nuclear burning of the hot CNO cycle in the envelope in addition to compressional and crustal heatings.
Given the accreted mass and $e$-folding time, our calculations can reproduce the observed light curve as a whole.
The two observations around 32 and 85 days locate significantly above our light curves.
The two flare observations might be due to increases in the quiescent accretion rate~\cite{2014ApJ...795..131H}
    or contamination due to the residual accretion~\cite{2015ApJ...809L..31D}.
 
Although we do not need to include an unknown shallow heat source different from the previous study~\cite{2015ApJ...809L..31D},
    we must discuss the light curve after 200 days.
As shown in Fig.~\ref{trho}, the hot CNO cycle expressed by the Eq.~(\ref{eq:CNO}) does not operate enough after 300 days
    because the formula can be applied for the temperature range of $0.2\leq T_9\leq 0.5$~\cite{1981ApJS...45..389W}.
Therefore, the effective temperature of our light curve decreases rather suddenly after 350 days.
To explore the agreement between theoretical light curve and observations,
    we need to use a nuclear reaction network coupled with the stellar evolutionary calculations;
    this gives changes in abundances with time.
However, this could lead to unstable nuclear burning and type I X-ray bursts.
This is beyond our present research because our aim in this investigation is to find unknown energy source.
To obtain the detailed light curve, we may further need to calculate X-ray bursts and elucidate how the bursts play a role against the quiescent luminosities.

Note that we have not included the effects of super-fluidity and viscous heating process because
    these have large uncertainties.
If these are included, we expect that the effective temperature will become higher because the super-fluidity suppresses the neutrino cooling rates,
    and the total heating rate increases because of the viscous heating.
If we decrease the initial mass accretion rate $dM / dt$ and choose its $e$-folding time $\tau$, 
    we may explain the observations of MAXI J0556-332 without the unknown shallow heat source.
Our aim is to investigate the physical source for the shallow heating of the nuclear burning.
Further discussions about the super-fluidity and viscous heating are beyond the scope of our present study.

In fact, the quiescent light curves of KS 1731-260 and MXB 1659-29~\cite{2009ApJ...698.1020B} require the heat source of approximately 1 MeV per accreted nucleon as the shallow heat source.
The relation between energy sources in the envelope and light curves depends on the history of the accretion and preceding X-ray bursts.
These issues are worth-while for studying the neutron star property.
It is significantly important to investigate whether light curves during quiescence eras can be reproduced with nuclear burning.
Furthermore, there remain problems such as the reheating event~\cite{2014ApJ...795..131H} and Urca cooling~\cite{2014Natur.505...62S} in the crust.
It may be very difficult to include all these phenomena and/or nuclear processes involving very uncertain nuclear processes.

\section*{Acknowledgment}
We would like to thank Kenzo Arai for useful comments.
This work was supported by JSPS KAKENHI Grant Numbers 24540278, 15K05083
and by a China Scholarship Council.

% can use a bibliography generated by BibTeX as a .bbl file
% BibTeX documentation can be easily obtained at:
% http://www.ctan.org/tex-archive/biblio/bibtex/contrib/doc/

\bibliographystyle{jpsj}
\bibliography{17310}
%
% once the .bbl file has been generated then place the text in your article.

%\begin{thebibliography}{100}
%
%\end{thebibliography}

%\appendix
%
%\section{Appendix head}

\end{document}